\documentclass[3p,authoryear,times]{elsarticle}

\usepackage{natbib}
\usepackage{amsmath,amssymb,bm}
\usepackage{graphicx}
\usepackage{caption}
\begin{document}

\begin{frontmatter}

\title{Multigame Effect in Finite Populations Induces \\
Strategy Linkage Between Two Games}

\author{Koh Hashimoto}
\ead{hasimoto@sacral.c.u-tokyo.ac.jp}
\address{Graduate School of Arts and Sciences,
The University of Tokyo, 3-8-1 Komaba,
Meguro, Tokyo 153-8902 Japan}

\begin{abstract}
Evolutionary game dynamics with two 2-strategy games in a finite population has been investigated in this study. Traditionally, frequency-dependent evolutionary dynamics are modeled by deterministic replicator dynamics under the assumption that the population size is infinite. However, in reality, population sizes are finite. Recently, stochastic processes in finite populations have been introduced into evolutionary games in order to study finite size effects in evolutionary game dynamics. However, most of these studies focus on populations playing only single games. In this study, we investigate a finite population with two games and show that a finite population playing two games tends to evolve toward a specific direction to form particular linkages between the strategies of the two games.
\end{abstract}


\end{frontmatter}

\section{Introduction}
Evolutionary game theory is a fundamental mathematical framework
that enables the investigation of evolution
in biological, social, and economic systems,
and has been successfully
applied to the study of the Darwinian process of natural selection
\citep{Lewontin:1961cd,Smith:1972tn,Smith:1973ut,MaynardSmith:1974bu,Smith:1982ww,Taylor:1978jg,Sugden:1986vq,Hofbauer:1998vd,Nowak:2006wr,Nowak:2004eb}.
The Darwinian process is an inherently frequency-dependent process.
The fitness of an individual is not only linked to environmental conditions
but also tightly coupled with the frequencies of its competitors.
Replicator dynamics, introduced by Taylor and Jonker \citep{Taylor:1978jg},
is a system of deterministic differential equations,
which model the frequency-dependent selection.
It is the most popular model for
the evolution of the frequencies of strategies in a population.
However, this model intrinsically assumes that
population sizes are infinite
and it fails to consider stochastic effects.

Recently, various frequency-dependent stochastic processes in finite populations
have been introduced
into evolutionary games in order to study the finite size effect in evolutionary
game dynamics \citep{Nowak:2004kv,Taylor:2004je,Fudenberg:2006um,Szabo:1998hl,Traulsen:2006cx,Szabo:2002cv}.
One such stochastic process is the frequency-dependent Moran process
\citep{Nowak:2004kv}.
This process is a stochastic birth-death process and comprises two procedures:
(1) birth, in which a player is chosen as a parent to reproduce with a probability
proportional to its fitness,
and its offspring has the same strategy as the parent,
and (2) death, in which the offspring replaces a randomly chosen individual.
Thus, the population size $N$ is strictly constant in both these procedures.
Another process is the local update process \citep{Traulsen:2005p127}.
In the local update process,
one individual is chosen randomly,
who compares his/her payoff to that of another randomly chosen individual,
and the probability of the former switching to the latter's strategy is based on
difference between their payoffs.
Repeating the process $N$-times is regarded as the unit time.
These models have enabled many analyses of evolutionary processes and
have provided considerable insight into the stochastic effects in evolutionary game
dynamics
\citep{Nowak:2004kv,Taylor:2004je,Fudenberg:2006um,Traulsen:2005p127,Ficici:2007p124,Ohtsuki:2007co}.
However, most of the previous studies have only
focused on a single game with a maximum of two to three strategies.
In most systems that are of interest to us,
we can see that players are playing many games simultaneously, such as biological
games in ecosystems and social games in human societies.
Such a situation where players play several games simultaneously is termed
{\it multigame} \citep{Hashimoto:2006p5}.
With an infinite population, the {\it multigame} effect has been investigated;
when the numbers of strategies of the games are more than two,
the fate of the frequencies of the strategies
in a single game may change dramatically with or without another game
in general \citep{Chamberland:2000fh,Hashimoto:2006p5}.
Even if one of the games has Evolutionary Stable Strategy (ESS) \citep{Smith:1982ww,Hofbauer:1998vd},
the ESS point may be destabilized \citep{Hashimoto:2006p5}.
However, when both games have two strategies, their fates coincide
with the fates of single games \citep{Cressman:2000kw}.
However, in a finite population, the manner in which
{\it multigame} influence the dynamics remains unclear.
In this article, we apply this motivation to the simplest case.

\section{Evolutionary game dynamics with two 2-strategy games}
In this section,
we investigate evolutionary game dynamics with two 2-strategy games
in a finite population.
Let us consider a population playing two games, game-$\alpha$ and game-$\beta$,
simultaneously. The reward matrices of the games are given by
\begin{align*}
  A = \begin{pmatrix} a_{11} & a_{12} \\
                      a_{21} & a_{22} \end{pmatrix}, \;
  B = \begin{pmatrix} b_{11} & b_{12} \\
                      b_{21} & b_{22} \end{pmatrix}, %\label{eq:AB}
\end{align*}
respectively.
The players can be divided into $2\times 2$ groups, with group
$(i,j)$ comprising players who play strategy-$i$ for game-$\alpha$
and strategy-$j$ for game-$\beta$.
We assume that a player's payoff for game-$\alpha$ and that for game-$\beta$
additively influence his/her payoff.
A player of strategy-$(i, j)$ playing against a player of
strategy-$(k, l)$ will be rewarded $a_{ik} + b_{jl}$.
This player obtains $a_{ik}$ through game-$\alpha$ and $b_{jl}$ through game-$\beta$.
Let $x_{ij}$ denote the frequency of players playing strategy-$(i,j)$
 ($\sum x_{ij}=1$).
Furthermore, the frequency of players playing strategy-$i$
in game-$\alpha$ is denoted by $y_i$  and
that of strategy-$j$ in game-$\beta$ by $z_j$:
\[ \bm{y}= \begin{pmatrix}  y_1 \\ y_2\end{pmatrix} =
\begin{pmatrix}  x_{11}+x_{12} \\ x_{21}+x_{22}\end{pmatrix},\;
\bm{z}= \begin{pmatrix}  z_1 \\ z_2\end{pmatrix} =
\begin{pmatrix}  x_{11}+x_{21} \\ x_{12}+x_{22}\end{pmatrix}.
\]
If every individual interacts with a representative sample of the population,
the expected payoff for an $(i,j)$-strategy player is determined by
$f_{ij}=\sum_{k,l}a_{ik}x_{kl}+b_{jl}x_{kl}$ and
the average payoff of the population is $\bar{f}=\sum x_{kl}f_{kl}$.
Using $\bm{y}$ and $\bm{z}$, these equations can be rewritten as follows:
\begin{align*}
f_{ij}=(A\bm{y})_i + (B\bm{z})_j,\quad \bar{f}={}^t\bm{y}A\bm{y}+{}^t\bm{z}B\bm{z}.
\end{align*}

In this article, we assume that
a coexistence equilibrium point exists in each game.
At a coexistence equilibrium point, all strategies obtain the same payoffs.
Let $\bm{p}={}^t(p,1-p)$ and $\bm{q}={}^t(q,1-q)$ denote
the coexistence equilibrium points in game-$\alpha$ and game-$\beta$,
respectively.
$p$ and $q$ are determined by
$p=\frac{a_{22}-a_{12}}{a_{11}-a_{12}-a_{21}+a_{22}}$ and
$q=\frac{b_{22}-b_{12}}{b_{11}-b_{12}-b_{21}+b_{22}}$, respectively.
The existence of the coexistence equilibrium points yields
$(a_{11}-a_{21})(a_{22}-a_{12}) > 0$ and $(b_{22}-b_{12})(b_{11}-b_{21}) > 0$.
Note that
in a {\it multigame} situation,
system states that satisfy the coexistence equilibria in both the games
are not a single point but rather points on a line determined by
$\bm{y}=\bm{p}$ and $\bm{z}=\bm{q}$.
Here, this line is termed $L$ (see Fig. \ref{fig:tetra1}).
Furthermore, we assume that
the equilibrium points are stable in both the games
and their stability are sufficiently strong.
The stability of the equilibrium points ensures
\begin{align*}
a_{11}-a_{21} < 0,&\quad a_{22}-a_{12} < 0, \\
b_{11}-b_{21} < 0,&\quad b_{22}-b_{12} < 0.
\end{align*}
This means that
the two games are Hawk and Dove games
which were initially introduced by J. Maynard Smith
\citep{Smith:1982ww}.

\subsection{Infinite-population model}
\begin{figure}[htb]
\begin{center}
 \includegraphics[width=2.8cm]{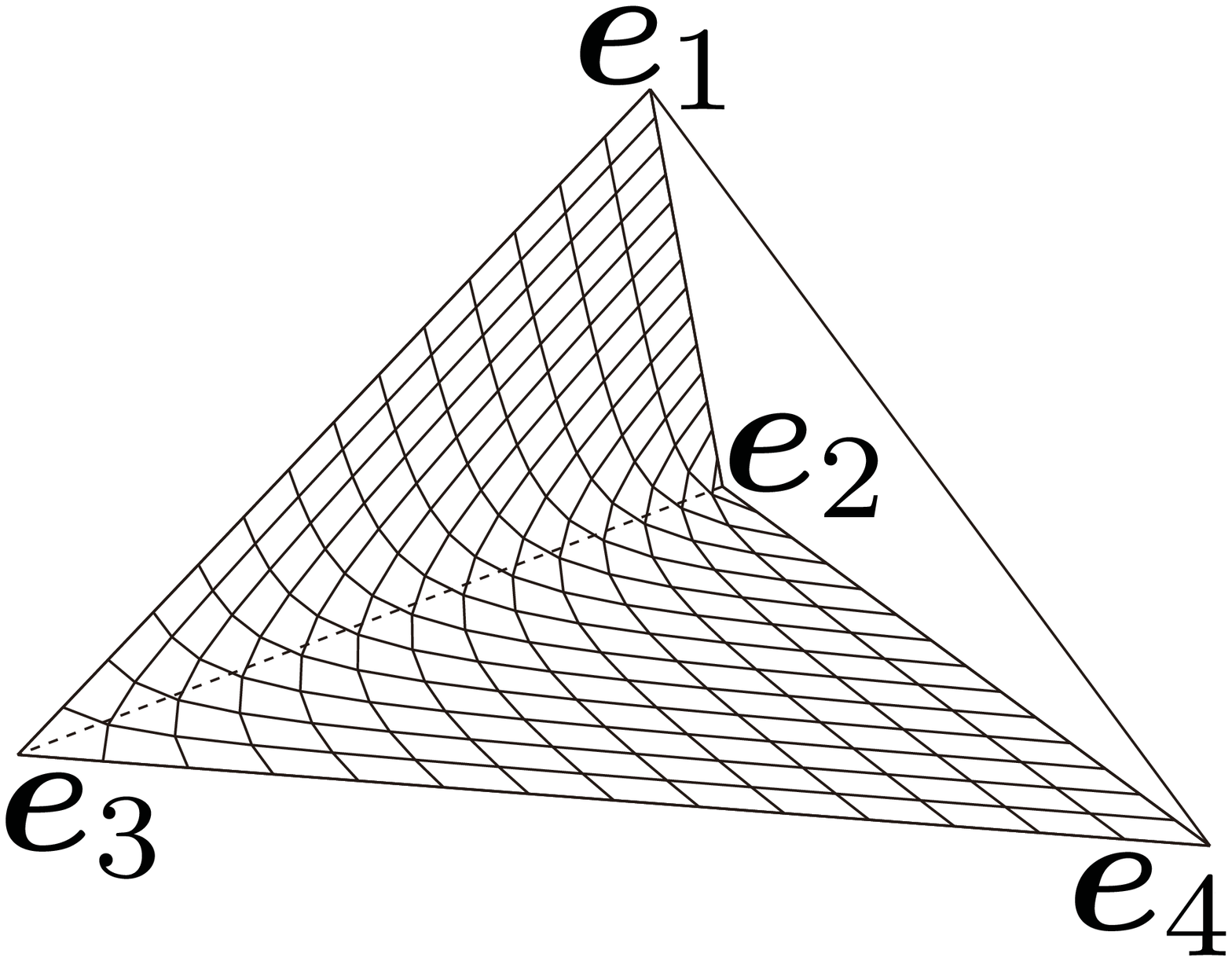}
 \includegraphics[width=2.8cm]{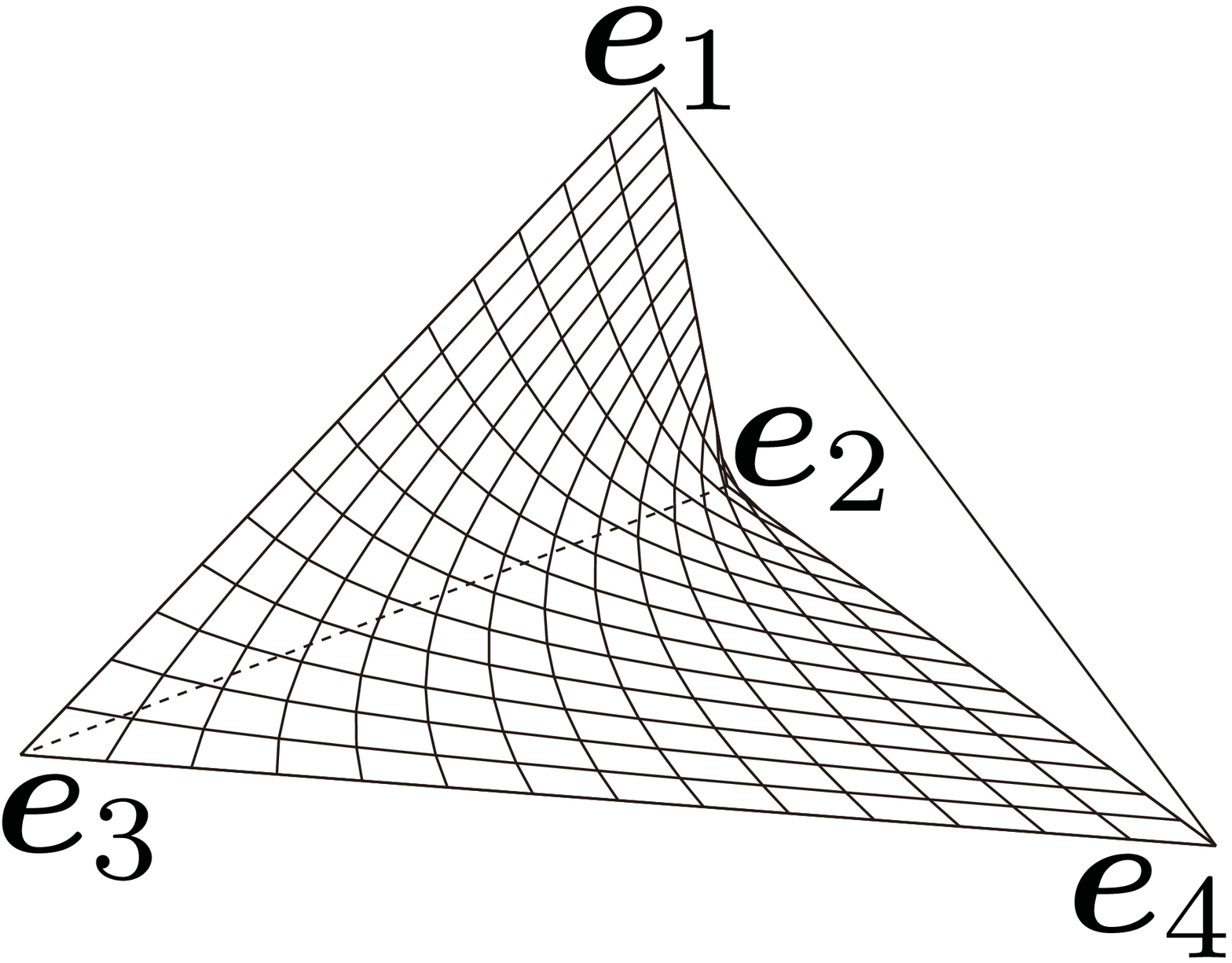}
 \includegraphics[width=2.8cm]{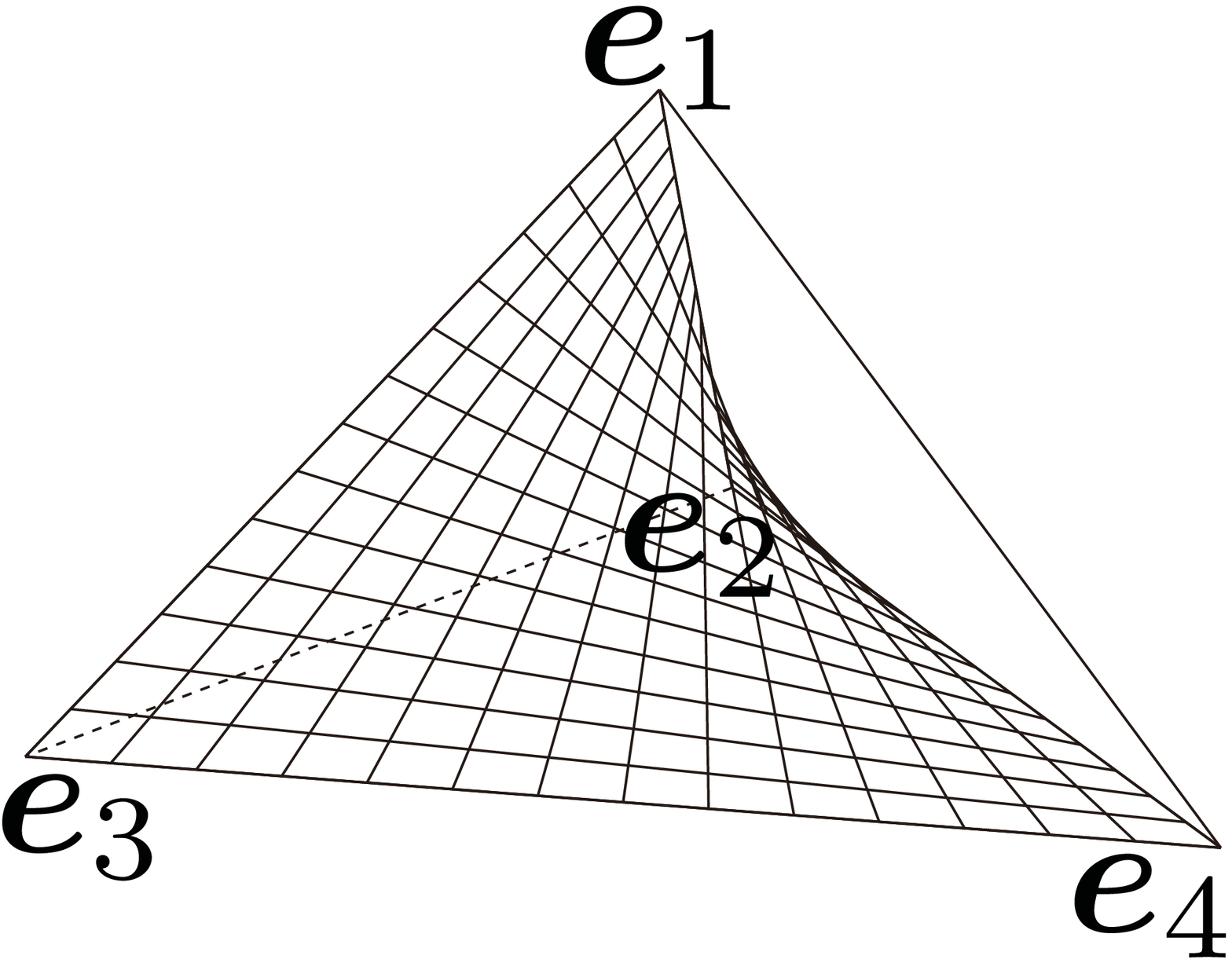}\\
(a)\hspace{2.4cm}(b)\hspace{2.4cm}(c)\\
 \includegraphics[width=2.8cm]{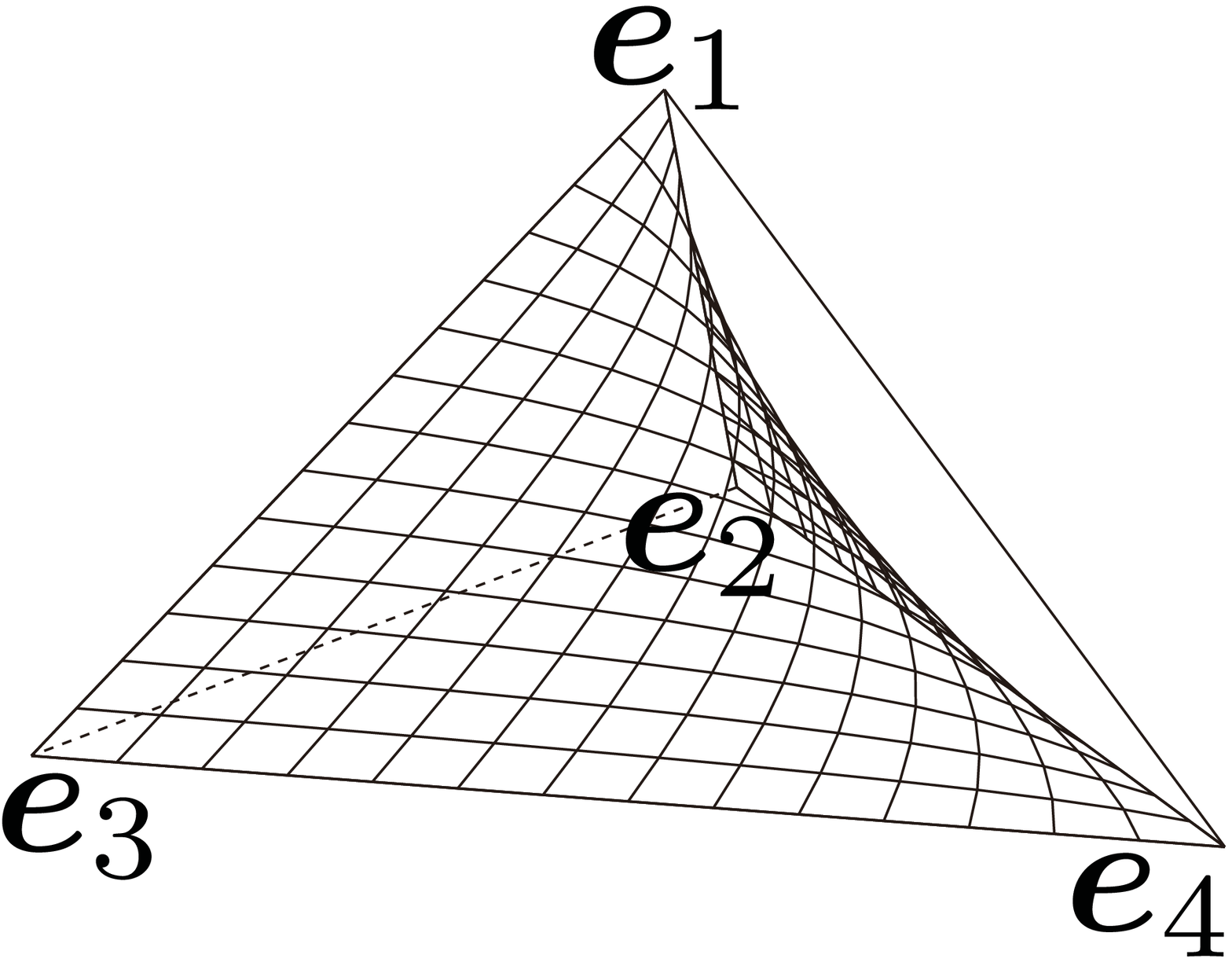}\hspace{1cm}
 \includegraphics[width=2.8cm]{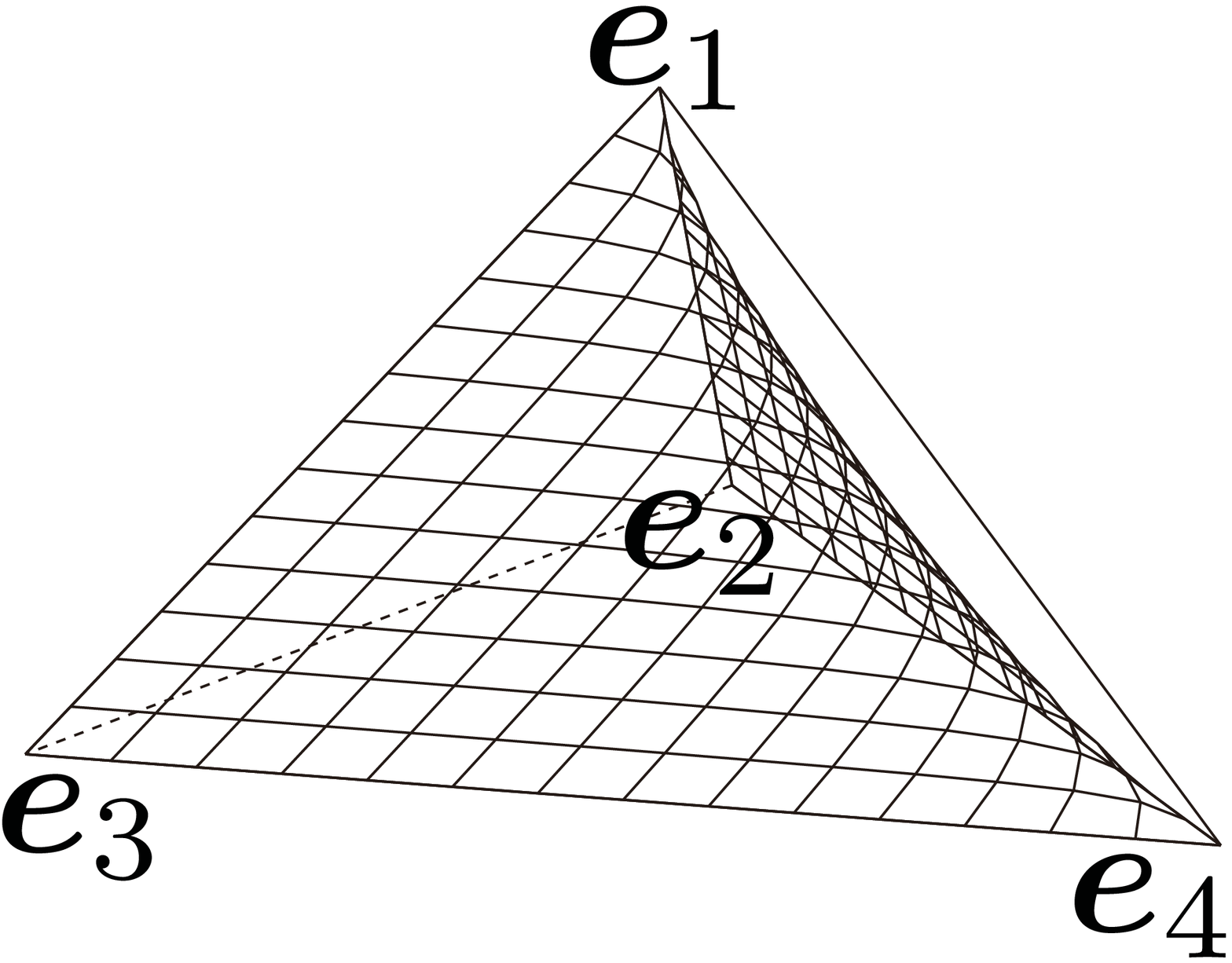}\\
(d)\hspace{3.6cm}(e)
\caption{Since $\sum x_{ij}=1$ and $x_{ij}\geq 0$ hold,
a system state can be represented by a point in a tetrahedron.
Five examples of invariant manifolds in the replicator equation
are plotted.
$\frac{x_{11}x_{22}}{x_{12}x_{21}}$ is
(a) $\frac{1}{25}$, (b) $\frac{1}{5}$, (c) $1$, (d) $5$, and (e) $25$.
The vertexes represent the system states that all players have the same strategy.
$\bm{e}_1$ represents one of the four vertexes: $(x_{11},x_{12},x_{21},x_{22})=(1,0,0,0)$.
$\bm{e}_2$,$\bm{e}_3$, and $\bm{e}_4$ represent
$(0,1,0,0)$, $(0,0,1,0)$, and $(0,0,0,1)$, respectively.
}
\label{fig:manifolds}
\end{center}
\end{figure}

\begin{figure}[htb]
\begin{center}
 \includegraphics[width=5.5cm]{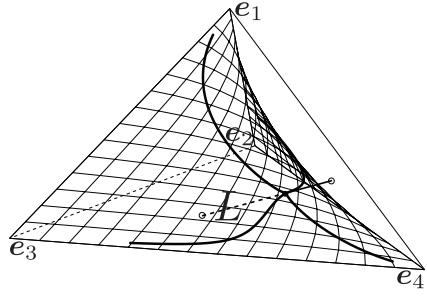}
\caption{A set of coexistence equilibrium points in this system forms a line
indicated by $L$.
Four orbits in a deterministic system with an infinite population are plotted.
All of these orbits start from points on an identical invariant manifold,
which is represented by a curved surface.
They move on the manifold and converge to the intersection point of
$L$ and the manifold.
}
\label{fig:tetra1}
\end{center}
\end{figure}

In an infinite population,
the replicator equation that corresponds to this {\it multigame} situation
is given by
\begin{align}
  \dot{x}_{ij}&=x_{ij}\left(f_{ij}-\bar{f}\right) \nonumber \\
&=x_{ij}\left\{(A\bm{y})_i+(B\bm{z})_j-
             {}^t\bm{y}A\bm{y}-{}^t\bm{z}B\bm{z}\right\}.  \label{eq:dotx}
\end{align}
Behaviors in this system are rather simple.
This differential equation leads
\begin{align*}
\frac{\mathrm{d}}{\mathrm{d}t}\frac{x_{11}x_{22}}{x_{12}x_{21}}=
\frac{x_{11}x_{22}}{x_{12}x_{21}}\left(f_{11}+f_{22}-f_{12}-f_{21}\right)=0.
\end{align*}
Thus, $\frac{x_{11}x_{22}}{x_{12}x_{21}}$ is constant in time evolution and
this implies that $\frac{x_{11}x_{22}}{x_{12}x_{21}}=(\mbox{const})$
forms an invariant manifold.
Some invariant manifolds are shown in Fig. \ref{fig:manifolds}.
Additionally, in Fig. \ref{fig:tetra1},
four orbits that follow Eq. (\ref{eq:dotx}) are plotted as examples.
An orbit starting at time $t=0$ from a point
%{\scriptsize
$\begin{pmatrix} x_{11}(0) & x_{12}(0)\\x_{21}(0) & x_{22}(0) \end{pmatrix}$%}
will move on an invariant manifold determined by
$\frac{x_{11}x_{22}}{x_{12}x_{21}}=\frac{x_{11}(0)x_{22}(0)}{x_{12}(0)x_{21}(0)}$
and will eventually converge to the intersection point of the
invariant manifold and $L$.
Every point on $L$ is a fixed point and
is stable to the transverse direction of $L$.
Furthermore, because every point on $L$ is a fixed point,
the stability to the direction of $L$ is neutral.

\subsection{Finite-population model}
In a finite population, stochastic effects by the demographic noise
constantly perturb the system state.
Selection pressures bring the system state to $L$,
however, stochastic effects prevent the system state from staying on $L$.
In this study, we are interested in determining
the motion of the system state in $L$'s direction.
From a naive intuition, it may seem to be a simple random walk
because the stability in this direction is neutral
in a system with an infinite population.
However, we find that
finite size effect breaks this neutrality and a flow emerges along $L$.
Consequently, the system state tends to evolve toward a specific
direction of $L$.

As mentioned above,
several processes are proposed for game dynamics in finite populations.
In this article, we concentrate on the local updating process proposed by
Traulsen et al. \citep{Traulsen:2005p127}.
In the local update process,
a player $b$ is chosen randomly and his/her payoff is compared to that of
another randomly chosen individual $a$.
The probability of player-$b$ switching his/her strategy to the player-$a$'s
strategy is given by
\begin{align*}
 \gamma(f_{(a)},f_{(b)})=
\frac{1}{2}+\frac{\omega}{2}(f_{(a)}-f_{(b)}),
\end{align*}
where $f_{(a)}$ and $f_{(b)}$ are the payoffs of player-$a$ and player-$b$,
respectively.
Furthermore, $\omega(\geq 0)$ measures the strength of the selection pressure.
Selection is weak when $\omega\ll 1$,
the process is dominated by random updating and
payoff differences have a negligible effect on the process.
For larger $\omega$, the selection intensity increases.
However, an upper limit is imposed on $\omega$ by the following requirement:
$0\leq \gamma\leq 1$.
The probability of selecting an $(i,j)$-strategy player as player-$a$
is $x_{ij}$
and the probability of selecting a $(k,l)$-strategy player as player-$b$
is $x_{kl}$.
Therefore, the probability, $P_{(k,l)}^{(i,j)}$,
that the number of $(i,j)$-strategy players increases by one
and that of $(k,l)$-strategy players decreases by one
in a single process is given by
\begin{align}
P_{(k,l)}^{(i,j)}=x_{ij}x_{kl}\gamma(f_{ij},f_{kl})=x_{ij}x_{kl}
\left\{\frac{1}{2}+
      \frac{\omega}{2}(f_{ij}-f_{kl})\right\}.
\label{eq:P}
\end{align}
The probability that the system state does not change is
$\displaystyle 1-\sum_{(i,j)\neq (k,l)} P_{(k,l)}^{(i,j)}$.

It is noteworthy that in the limit of $N\rightarrow\infty$,
this process represents a replicator equation.
The expected change of $x_{ij}$ in a single process is as follows:
\begin{align*}
\left<\Delta x_{ij}\right>&=\frac{1}{N}\sum_{(k,l)} P^{(i,j)}_{(k,l)} - P^{(k,l)}_{(i,j)}
=\frac{\omega}{N} x_{ij}\left(f_{ij}- \bar{f}\right).
\end{align*}
Because the process is iterated $N$ times in a unit time,
a single process takes time $\Delta t = \frac{1}{N}$.
In the limit of $N\rightarrow\infty$, the replicator equation is derived as
\begin{align*}
\dot{x}_{ij} &= \lim_{N\rightarrow\infty} \frac{\Delta x_{ij}}{\Delta t}
=\omega x_{ij}\left(f_{ij}- \bar{f}\right).
\end{align*}
Therefore, the stochastic process indeed implements
a replicator equation in a finite population with stochasticity of demographic noise.

\begin{figure}[htb]
\begin{center}
 \includegraphics[width=5.5cm]{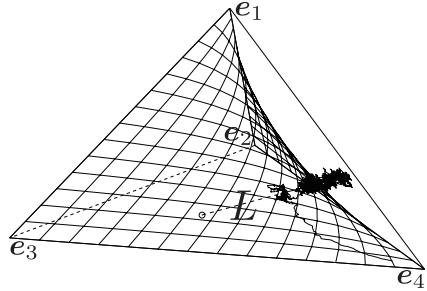}
\caption{A trajectory in a stochastic system with a finite population is
plotted.
The system state is brought to $L$ due to the selection pressure
in the same way as the infinite-population model.
However, in contrast to the case of an infinite population,
stochastic effects prevent the system state from staying on $L$,
and it fluctuates around $L$.
Eventually, the system state reaches one of the boundaries of $L$.
}
\label{fig:tetra2}
\end{center}
\end{figure}

In the process,
the system state ultimately reaches
one of the four homogeneous states
(i.e., the states that all players have the same strategy)
after long transient,
and these four states correspond to the four vertexes of the tetrahedron.
However, the time taken to reach such a homogeneous state is extremely long,
especially in a large population,
because of the stability of coexistence equilibrium points
\citep{Antal:2006ji}.
In contrast, extinction of a single strategy resulting from a random walk along
$L$ occurs in a short time-scale.
Similar to the infinite-population case,
first, the system state is immediately brought to the vicinity of $L$ from
its initial point by the selection pressure roughly
along the manifold that the initial state is located.
Subsequently, the system state fluctuates around $L$ because of the stochastic effects.
Eventually, one of the strategies become extinct,
implying that the system state has reached one of the boundaries of $L$.
Strictly speaking, this is inaccurate. In fact, the system state reaches
a point on a boundary plane when a single strategy is extinct and this location is
not necessarily a boundary of $L$ but a point close to it.
Following this, the system state fluctuates on the boundary plane
around the intersection point of $L$ and the boundary plane.
In Fig. \ref{fig:tetra2}, an evolutionary path from an initial state
to a state in which a single strategy is eliminated is plotted.
Finally, in the limit of time evolution, two strategies will be eliminated,
that is, the population becomes homogeneous.
This is the absorbing state of the system
and it takes an enormously long time because of the stability of the coexistence
equilibrium.
Therefore, the boundaries of $L$ can be considered as absorbing states for a
short time-scale evolution.
As mentioned above,
we are not interested in determining how the homogeneous states are reached
but rather how the boundaries of $L$ are reached.
This rather literary illustration of the system behavior raises following questions:
Is the movement of the system state along $L$ just a simple random walk?
If not, which absorbing state has a higher likelihood of being observed?
This article is aimed to answer these questions.

\subsection{Variable transformation}
\begin{figure}[htb]
\begin{center}
 \includegraphics[width=6cm]{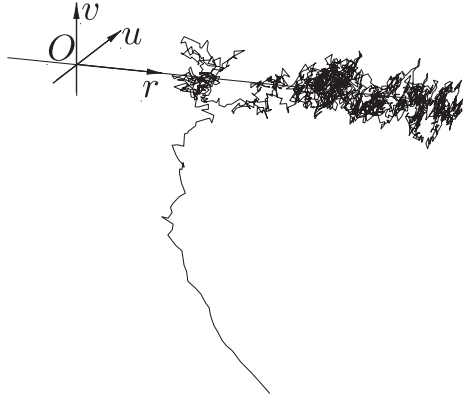}
\caption{
A sample trajectory in $(r,u,v)$-coordinate.
$r$-axis is identical to the line $L$ indicated in Fig. \ref{fig:tetra2}.
The trajectory makes its way toward the $r$-axis and subsequently
fluctuates around it.
Eventually, it reaches one of the boundaries of $L$.
}
\label{fig:ruv}
\end{center}
\end{figure}

To investigate the behavior of the system
we here adopt a variable transformation
$\begin{pmatrix} x_{11} & x_{12} \\ x_{21} & x_{22} \end{pmatrix}
\mapsto (r,u,v)$:
\begin{align}
  &\begin{pmatrix} x_{11} & x_{12} \\ x_{21} & x_{22} \end{pmatrix}
  =
\begin{pmatrix}p q  & p (1-q) \\ (1-p)q  & (1-p)(1-q)\end{pmatrix}% \nonumber \\
%&\quad
+ \frac{r}{2} \begin{pmatrix}  1 & -1  \\  -1 & 1 \end{pmatrix}
+ \frac{u}{2} \begin{pmatrix}  1 & 1  \\  -1 & -1 \end{pmatrix}
+ \frac{v}{2} \begin{pmatrix}  1 & -1  \\  1 & -1 \end{pmatrix}. \label{eq:ruv}
\end{align}
Note that the origin of $(r,u,v)$-coordinate depends on $\bm{p}$ and $\bm{q}$.
Since Eq. (\ref{eq:ruv}) leads
% $\bm{y} = \bm{p} + u {\scriptsize \begin{pmatrix}1 \\ -1 \end{pmatrix}}$ and
% $\bm{z} = \bm{q} + v {\scriptsize \begin{pmatrix}1 \\ -1 \end{pmatrix}}$,
$\bm{y} = \bm{p} + u \begin{pmatrix}1 \\ -1 \end{pmatrix}$ and
$\bm{z} = \bm{q} + v \begin{pmatrix}1 \\ -1 \end{pmatrix}$,
$u$ and $v$ represents the deviations from equilibria in game-$\alpha$ and
game-$\beta$, respectively.
$L$ is represented by $u=v=0$. Therefore, $r$-axis is identical to $L$.
A sample trajectory in $(r,u,v)$-coordinate is plotted in Fig. \ref{fig:ruv}.
A Markov process on $(r,u,v)$-space is led from the transition probabilities (Eq. (\ref{eq:P})) and this variable transformation.
For an example, with a probability of $P^{(1,1)}_{(1,2)}$,
a $(1,2)$-strategy player is replaced by $(1,1)$-strategy player,
then $r$ and $v$ increase by $1/N$ and $u$ remains unchanged.
Another example,
a $(2,2)$-strategy player is replaced by $(1,1)$-strategy player
with a probability of $P^{(1,1)}_{(2,2)}$,
then $u$ and $v$ increase by $1/N$ and $r$ remains unchanged.

\subsection{Langevin equations}
To analyze the dynamics of a large but finite population,
it is effective to approximate the dynamics using Langevin equations
\citep{Helbing:1996bw,Traulsen:2005p127,Traulsen:2006cx,Traulsen:2012ww}.
The objective of this study is to determine the time evolution of $r$;
the first step to this is to derive the Langevin equations for
$(r,u,v)$ by approximating the Markov process, assuming
that the population size $N$ is sufficiently large.

The expected changes of $r$, $u$ and $v$ in a single process
are respectively given as
\begin{align*}
\left<\Delta r\right> &= \frac{1}{N}g_r(r,u,v)=
       \frac{1}{N}\sum_{\substack{i \\k\neq l}}P_{(k,l)}^{(i,i)}-P_{(i,i)}^{(k,l)}, \\
\left<\Delta u\right> &= \frac{1}{N}g_u(r,u,v)=
          \frac{1}{N}\sum_{k,l}P_{(2,l)}^{(1,k)}-P_{(1,k)}^{(2,l)}, \\
\left<\Delta v\right> &= \frac{1}{N}g_v(r,u,v)=
          \frac{1}{N}\sum_{k,l}P_{(l,2)}^{(k,1)}-P_{(k,1)}^{(l,2)}.
\end{align*}
These are used for the drift terms in the Langevin equations.
Based on a simple calculation using Eq. (\ref{eq:P}) and the variable transformation Eq. (\ref{eq:ruv}),
$g_r$, $g_u$, and $g_v$ are given by
\begin{align}
g_r(r,u,v)
   &=\omega\left[
     \varPhi u\left\{-p (1-p)(1-2p)- ru
               +\frac{r}{2}(1-2p) \right.%\right. \nonumber \\
%     &\quad\quad\quad\quad\quad\left.
               -\frac{u}{2}(1-2p)(1-2q)+\frac{v}{2}\right\}\nonumber \\
   &\quad\quad
    +\varPsi v \left\{-q (1-q)(1-2q)- rv
                +\frac{r}{2}(1-2q)%\right.\nonumber \\
%     &\quad\quad\quad\quad\quad\left.
               \left.-\frac{v}{2}(1-2p)(1-2q)+\frac{u}{2}\right\}
            \right] \label{eq:gr}\\
g_u(r,u,v)
      &=\omega\left[\varPhi u (p+u) (1-p-u) %\nonumber \\
%            &\left.\;
              +\varPsi v \left\{\frac{r}{2}-uv+\frac{u}{2}(1-2q)+\frac{v}{2}(1-2p)
        \right\}\right] \label{eq:gu}\\
g_v(r,u,v)
      &=\omega\left[\varPsi v (q+v) (1-q-v) %\nonumber \\
%            &\left.\;
              +\varPhi u \left\{\frac{r}{2}-uv+\frac{u}{2}(1-2q)+\frac{v}{2}(1-2p)
        \right\}\right]\label{eq:gv}
\end{align}
respectively, where
$\varPhi = a_{11}+a_{22}-a_{12}-a_{21}, \varPsi = b_{11}+b_{22}-b_{12}-b_{21}$.
Note that the stability of the equilibrium points in the two games yields
$\varPhi < 0, \varPsi < 0$.
Moreover,
$\left<(\Delta r)^2\right>$, $\left<(\Delta u)^2\right>$,
$\left<(\Delta v)^2\right>$, $\left<\Delta r \Delta u\right>$,
$\left<\Delta r \Delta v\right>$, $\left<\Delta u \Delta v\right>$
are also respectively given as
\begin{align*}
\left<(\Delta r)^2\right> &=\frac{1}{N^2}h_{r r}(r,u,v)
%         =\frac{1}{N^2}\sum_{\ast\star} \mu_{+\ast\star}+\mu_{-\ast\star}, \\
    =\frac{1}{N^2}\sum_{\substack{i \\k\neq l}}P_{(k,l)}^{(i,i)}+P_{(i,i)}^{(k,l)}, \\
\left<(\Delta u)^2\right> &=\frac{1}{N^2}h_{u u}(r,u,v)
%          =\frac{1}{N^2}\sum_{\ast\star} \mu_{\ast+\star}+\mu_{\ast-\star}, \\
          =\frac{1}{N^2}\sum_{k,l}P_{(2,l)}^{(1,k)}+P_{(1,k)}^{(2,l)}, \\
\left<(\Delta v)^2\right> &=\frac{1}{N^2}h_{v v}(r,u,v)
%          =\frac{1}{N^2}\sum_{\ast\star} \mu_{\ast\star+}+\mu_{\ast\star-}, \\
          =\frac{1}{N^2}\sum_{k,l}P_{(l,2)}^{(k,1)}+P_{(k,1)}^{(l,2)}, \\
\left<\Delta r \Delta u\right>&=\frac{1}{N^2}h_{r u}(r,u,v)
%          =\frac{1}{N^2}\sum_{\ast} \mu_{++\ast}+\mu_{--\ast}-\mu_{+-\ast}-\mu_{-+\ast},\\
          =\frac{1}{N^2}\sum_{k\neq l}P_{(l,1)}^{(k,1)}-P_{(l,2)}^{(k,2)}, \\
%&          =\frac{1}{N^2}P_{21}^{11}+P_{11}^{21}-P_{22}^{12}-P_{12}^{22}, \\
\left<\Delta r \Delta v\right>&=\frac{1}{N^2}h_{r v}(r,u,v)
          =\frac{1}{N^2}\sum_{k\neq l}P_{(1,l)}^{(1,k)}-P_{(2,l)}^{(2,k)}, \\
%&          =\frac{1}{N^2}P_{12}^{11}+P_{11}^{12}-P_{22}^{21}-P_{21}^{22}, \\
\left<\Delta u \Delta v\right>&=\frac{1}{N^2}h_{u v}(r,u,v)
          =\frac{1}{N^2}\sum_{k\neq l}P_{(l,l)}^{(k,k)}-P_{(k,l)}^{(l,k)}.
%&          =\frac{1}{N^2}P_{22}^{11}+P_{11}^{22}-P_{21}^{12}-P_{12}^{21}.
\end{align*}
These are used for the diffusion terms in the Langevin equations.
Remember that the process is repeated $N$ times in a unit time.
Assuming that $N$ is sufficiently large,
the central limit theorem yields the Langevin equations as follows:
\begin{align}
  \dot{r} &= g_r(r,u,v) + \frac{1}{\sqrt{N}}\zeta_r(t), \label{eq:langevin1r}\\
  \dot{u} &= g_u(r,u,v) + \frac{1}{\sqrt{N}}\zeta_u(t), \label{eq:langevin1u}\\
  \dot{v} &= g_v(r,u,v) + \frac{1}{\sqrt{N}}\zeta_v(t). \label{eq:langevin1v}.
\end{align}
Here, $\zeta_r(t)$, $\zeta_u(t)$ and $\zeta_v(t)$ denote Gaussian white noise
with
\begin{align*}
% \left<\zeta_r(t)\right>=0 &
% \left<\zeta_v(t)\right>=0 &
% \left<\zeta_v(t)\right>=0 \\
\left<\zeta_r(t)\right>&=\left<\zeta_u(t)\right>=\left<\zeta_v(t)\right>=0, \\
\left<\zeta_r(t)\zeta_r(t')\right>&=\left(h_{r r}-g_r^2\right) \delta(t-t'), \\
\left<\zeta_u(t)\zeta_u(t')\right>&=\left(h_{u u}-g_u^2\right) \delta(t-t'), \\
\left<\zeta_v(t)\zeta_v(t')\right>&=\left(h_{v v}-g_v^2\right) \delta(t-t'), \\
\left<\zeta_r(t)\zeta_u(t')\right>&=\left(h_{r u}-g_rg_u\right) \delta(t-t'), \\
\left<\zeta_r(t)\zeta_v(t')\right>&=\left(h_{r v}-g_rg_v\right) \delta(t-t'), \\
\left<\zeta_u(t)\zeta_v(t')\right>&=\left(h_{u v}-g_ug_v\right) \delta(t-t').
\end{align*}
These noises are demographic stochasticity and
should be interpreted in Ito's sense.
Assuming that the system size $N$ is sufficiently large leads that
the deviations of $u$ and $v$ are sufficiently small.
Under this assumption, we disregard the effects of the deviations of $u$ and $v$ on the noise terms.
For an example,
$\left<\zeta_r(t)\zeta_r(t)\right>=h_{r r}(r,u,v)-g_r^2(r,u,v)$ can be approximated by
$h_{r r}(r,0,0)-g_r^2(r,0,0)$. The other correlations of the noise
terms can be approximated in the same manner. A simple calculation yields the
following equations:
\begin{align}
h_{r r}-g_r^2&\approx h_{r r}(r,0,0)-g_r^2(r,0,0) \nonumber \\
&=\left\{p q+(1-p)(1-q)+r\right\}%\times  \nonumber \\
%&\quad\quad
\left\{p(1-q)+(1-p)q-r\right\} \label{eq:hrr}\\
h_{u u}-g_u^2&\approx h_{u u}(r,0,0)-g_u^2(r,0,0)=p(1-p) \label{eq:huu}\\
h_{v v}-g_v^2&\approx h_{v v}(r,0,0)-g_v^2(r,0,0)=q(1-q) \label{eq:hvv}\\
h_{u v}-g_ug_v&\approx h_{u v}(r,0,0)-g_u(r,0,0)g_v(r,0,0)=\frac{r}{2} \label{eq:huv} \\
h_{r u}-g_rg_u&\approx h_{r u}(r,0,0)-g_r(r,0,0)g_u(r,0,0)%\nonumber\\&
=-(1-2p)q(1-q) \label{eq:hru} \\
h_{r v}-g_rg_v&\approx h_{r v}(r,0,0)-g_r(r,0,0)g_v(r,0,0)%\nonumber\\&
=-(1-2q)p(1-p) \label{eq:hrv}
\end{align}
Thus, the correlations of noise $\zeta_u$ and $\zeta_v$ are approximated by
\begin{align*}
\left<\zeta_u(t)\zeta_u(t')\right>&=p(1-p) \delta(t-t'), \\
\left<\zeta_v(t)\zeta_v(t')\right>&=q(1-q) \delta(t-t'), \\
\left<\zeta_u(t)\zeta_v(t')\right>&=\frac{r}{2} \delta(t-t'),
\end{align*}
respectively.
This approximation and linearizing
Eqs. (\ref{eq:langevin1u}) and (\ref{eq:langevin1v}) yield
\begin{align}
%  \frac{\mathrm{d}}{\mathrm{d}t}\begin{pmatrix} u \\ v \end{pmatrix} =
  \begin{pmatrix} \dot{u} \\ \dot{v} \end{pmatrix} =
J\begin{pmatrix} u \\ v \end{pmatrix}
+\frac{1}{\sqrt{N}} \begin{pmatrix} \zeta_u(t) \\ \zeta_v(t) \end{pmatrix}
%+\frac{1}{\sqrt{N}} \begin{pmatrix} \xi_u(t) \\ \xi_v(t) \end{pmatrix}
\label{eq:langevin2uv}
\end{align}
where $J$ is the Jacobi matrix of Eqs. (\ref{eq:langevin1u}) and (\ref{eq:langevin1v}) at $(u,v)=(0,0)$:
\begin{align*}
  J=
  \begin{pmatrix}
    \frac{\partial g_u}{\partial u} & \frac{\partial g_u}{\partial v} \\
    \frac{\partial g_v}{\partial u} & \frac{\partial g_v}{\partial v}
  \end{pmatrix}=
\omega\begin{pmatrix}
\varPhi p(1-p) & \varPsi \frac{r}{2} \\ \varPhi \frac{r}{2} & \varPsi q (1-q)\end{pmatrix}.
\end{align*}
Because $N$ is sufficiently large, we can assume that
the motion of $r$ is rather slow and
the time taken to reach one of the boundaries of $L$
is sufficiently long.
Thus, the asymptotic solution of the linear Langevin equation
Eq. (\ref{eq:langevin2uv}) is given as
\begin{align}
  \begin{pmatrix}u(t) \\ v(t)\end{pmatrix} = \frac{1}{\sqrt{N}}\int_{-\infty}^t
e^{J(t-s)}
  \begin{pmatrix}\xi_u(s) \\ \xi_v(s)\end{pmatrix} \mathrm{d} s.
\label{eq:uvint}
\end{align}
Based on a simple algebra, it is shown that
$u$ and $v$ obey the two-dimensional normal distribution with zero mean:
\begin{align*}
  \frac{1}{2\pi\sqrt{|\varGamma|}}\exp\left[-\frac{1}{2}(u,v)\varGamma^{-1}
    \begin{pmatrix}u \\ v\end{pmatrix}\right]
\end{align*}
where $\varGamma$ is the covariance matrix of $u$ and $v$:
\begin{align}
  \varGamma= \left<\begin{pmatrix}u^2 & uv \\ uv & v^2\end{pmatrix}\right>
        = -\frac{1}{2N\omega}
  \begin{pmatrix}\frac{1}{\varPhi} & 0 \\ 0 & \frac{1}{\varPsi}\end{pmatrix}.
  \label{eq:Gamma}
\end{align}

\begin{figure}[htb]
\begin{center}
 \includegraphics[width=7.5cm]{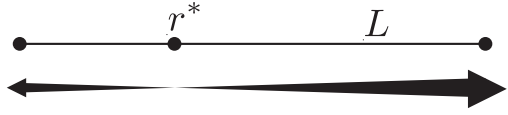}\\
\caption{A schematic view of the drift term of Eq. (\ref{eq:dotr}).
}
\label{fig:r}
\end{center}
\end{figure}

For the next step,
we approximate $g_r(r,u,v)$ by $\left<g_r(r,u,v)\right>_{u,v}$
in order to exclude $u$ and $v$ from the Langevin equation of $r$
(Eq. (\ref{eq:langevin1r})).
By using Eq. (\ref{eq:Gamma}), we obtain
\begin{align*}
\left<g_r(r,u,v)\right>_{u,v}
=\frac{1}{N}\left\{r+\frac{1}{2}(1-2p)(1-2q)\right\}.
\end{align*}
Finally, Eq. (\ref{eq:langevin1r}) can be approximated  by
\begin{align}
  \left<\dot{r}\right>_{u,v}=\frac{1}{N}\left\{r+\frac{1}{2}(1-2p)(1-2q)\right\}
         +\frac{1}{\sqrt{N}}\zeta_r(t) \label{eq:dotr}
\end{align}
where $\zeta_r(t)$ denotes Gaussian white noise with
\begin{align*}
\left<\zeta_r(t)\right>&=0, \\
\left<\zeta_r(t)\zeta_r(t')\right>&=
\left\{p q+(1-p)(1-q)+r\right\}%\times \\
%&\quad\quad\quad\quad
\left\{p(1-q)+(1-p)q-r\right\}\delta(t-t').
\end{align*}
In contrast to the case of an infinite population,
Eq. (\ref{eq:dotr}) indicates that the neutrality on $L$ vanishes and
the motion of $r$ is not a simple random walk.
The drift term of Eq. (\ref{eq:dotr}) is positive (negative)
when $r$ is larger (smaller) than $r^*=-\frac{1}{2}(1-2p)(1-2q)$.
Thus, the equilibrium point $r^*$ is {\it unstable} in Eq. (\ref{eq:dotr})
(see Fig. \ref{fig:r}).
The order of the drift term is $O(1/N)$,
whereas that of the diffusion term is $O(1/\sqrt{N})$.
Therefore, the {\it flow} by the drift term
is obscured by the diffusion
when the population size is large.
However, if the population size is not large,
distinct trends of behaviors can be observed
as demonstrated by the numerical simulations in Section 3.

Note that from Eq. (\ref{eq:gr}),
it is obvious that
$\left<\Delta r\right>$ is always zero when $(u,v)=(0,0)$
(i.e., the system state is precisely on $L$)
regardless of the value of $r$.
Therefore,
when the system state is precisely on $L$,
the expected motion of $r$ is neutral.
However, constantly occurring perturbation by finite size effect
causes $\left<u^2\right>$ and $\left<v^2\right>$ to be non-zero,
making {\it flow} presented in Fig. \ref{fig:r} in
the expected motion of $r$.
In this manner, emergence of such {\it flow} is
a somewhat indirect effect of finiteness of the system size.

\subsection{Strategy linkage between the two games}
\begin{figure}[htb]
\begin{center}
 \includegraphics[width=8cm]{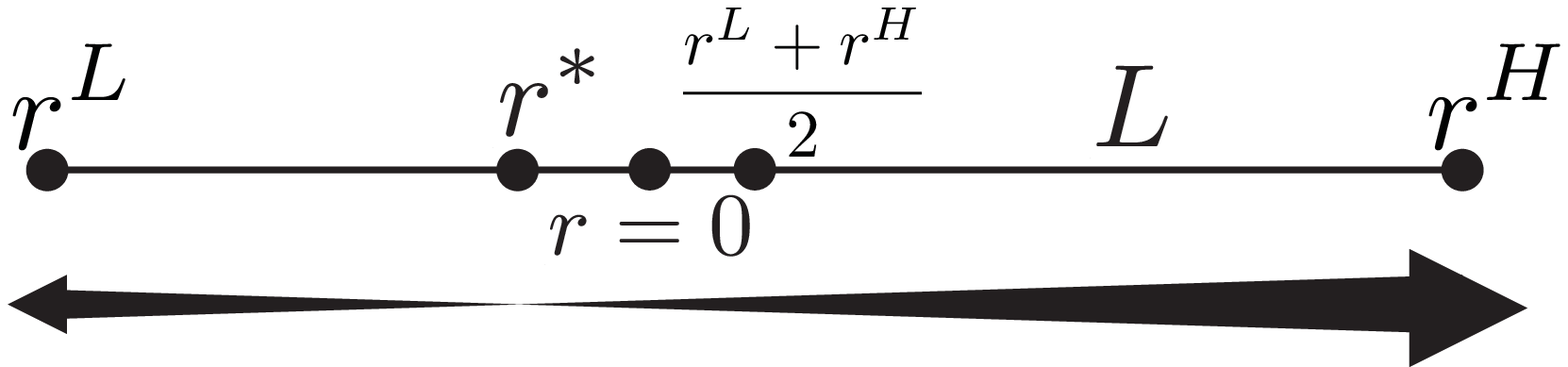} \\
(a) \\
 \includegraphics[width=8cm]{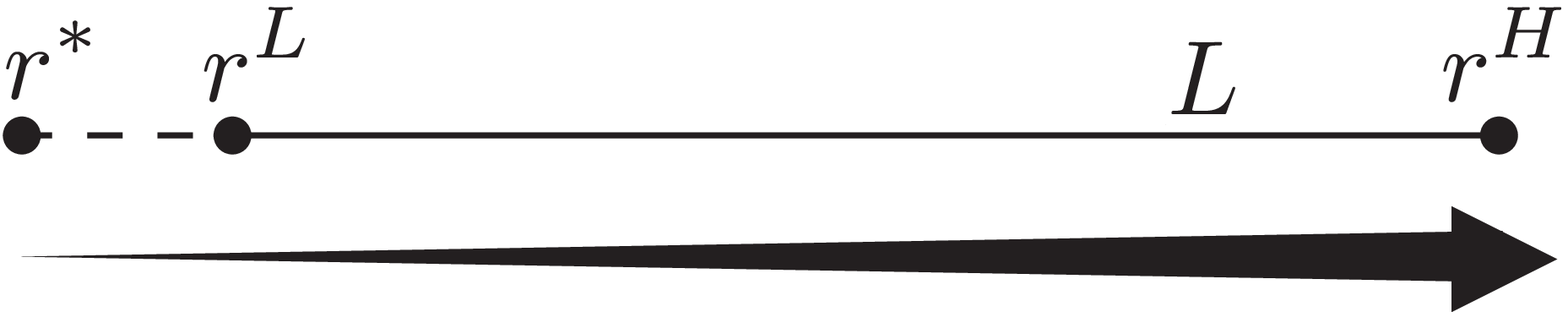} \\
(b) \\
\caption{Schematic views of the drift term in Eq. (\ref{eq:dotr}).
Since $p,q \leq \frac{1}{2}$,
$r^* \leq 0 \leq \frac{r^L+r^H}{2}$ is always satisfied.
(a) When $p+q \geq \frac{1}{2}$, $r^*$ is between
$r^L$ and $r=0$.
(b) Otherwise, $r^*$ is under $r^L$.
}
\label{fig:r12}
\end{center}
\end{figure}

On $L$, $r$ measures the linkage between the strategies of the two games.
$x_{ij}=y_iz_j$ holds if $r=0$, that is, there is no linkage between the strategies of the two games.
When $r$ is positively larger,
a player playing strategy-1 (strategy-2) in game-$\alpha$
also plays strategy-1 (strategy-2) in game-$\beta$ with a higher probability
and vice versa.
In such a case, the strategies of the two games are positively linked in the population.
On the other hand, when $r$ is negatively larger,
a player playing strategy-1 (strategy-2) in game-$\alpha$
plays strategy-2 (strategy-1) in game-$\beta$ with higher probability
and vice versa; in such a case,
the strategies of the two games are negatively linked.
Because the labels of the strategies can be assigned arbitrarily,
we can assume that $p,q \leq \frac{1}{2}$ without the loss of generality.
This assumption means that in each game,
the minor strategy at the equilibrium point is labeled strategy-1 and
the major one is labeled strategy-2.
Here, we show that the system state tends to have positive linkage under this condition.
We can also assume that $p \leq q$.
Even if $p > q$, game-$\alpha$ and game-$\beta$ just need to be relabeled
to satisfy this assumption.
These assumptions yield the following equations: $p q\leq p(1-q) \leq (1-p)q\leq (1-p)(1-q)$.
Therefore, non-negativity of $x_{ij}$ indicates
that the boundaries of $L$ are
%{\scriptsize $\begin{pmatrix} 0 & p \\ q & 1-p-q\end{pmatrix}$}
$\begin{pmatrix} 0 & p \\ q & 1-p-q\end{pmatrix}$
and
%{\scriptsize $\begin{pmatrix} p & 0 \\ q-p & 1-q\end{pmatrix}$}.
$\begin{pmatrix} p & 0 \\ q-p & 1-q\end{pmatrix}$.
Thus, the range of $r$ is given by
\begin{align*}
%  -2p q(=r^L) \leq r \leq 2p(1-q)(=r^H).
  r^L \leq r \leq r^H,\; \left(r^L=-2p q,\; r^H=2p(1-q)\right).
\end{align*}
Since $p,q \leq \frac{1}{2}$,
$r^*$ is always non-positive and
the midpoint of $L$ is always non-negative
%(i.e. $r^* \leq 0 \leq \frac{r^L+r^H}{2}=p(1-2q)$).
(i.e., $r^* \leq 0 \leq \left(r^L+r^H\right)/2=p(1-2q)$).
Thus, the position of $r^*$ on $L$ can typically be depicted as Fig. \ref{fig:r12}(a).
% Furthermore, if $(1-p)(1-q) < \frac{1}{2}$ is satisfied,
Furthermore, if $p+q < \frac{1}{2}$ is satisfied,
$r^*$ is lesser than $r^L$ and
in such a case, the drift term in Eq. (\ref{eq:dotr}) is
always set to be positive regardless of the value of $r$,
as presented in Fig. \ref{fig:r12}(b).
Such an obvious asymmetry of {\it flow} introduces a particular trend in the system.
If the initial state is chosen randomly,
it is clear that the system state goes to the higher boundary of $L$ with
a higher probability than the lower boundary.
The population tends to have {\it minor-minor and major-major linkages}
between the strategies of the two games.

Additionally, let us consider the rare mutation of strategies.
%Additionally, let us introduce rare mutation into this system.
Mutation restores extinct strategies and makes the system ergodic.
Therefore, the system state can be observed to making round trips along $L$ repeatedly.
Obviously, states around
%{\scriptsize $\begin{pmatrix} p & 0 \\ q-p & 1-q\end{pmatrix}$}
$\begin{pmatrix} p & 0 \\ q-p & 1-q\end{pmatrix}$
are observed more often than
$\begin{pmatrix} 0 & p \\ q & 1-p-q\end{pmatrix}$.
This indicates that the population with positive linkage is observed
more often than that with negative linkage.
Although it is difficult to analytically derive the frequencies of
the periods to stay around the lower and higher boundaries, we will
confirm this bias with a numerical simulation in Section 3.

\section{Numerical simulations}
Here, we demonstrate the tendency of the motion of $r$ with several numerical simulations.
\subsection{Round trips along $L$}
\begin{figure}[htb]
\begin{center}
 \includegraphics[width=7cm]{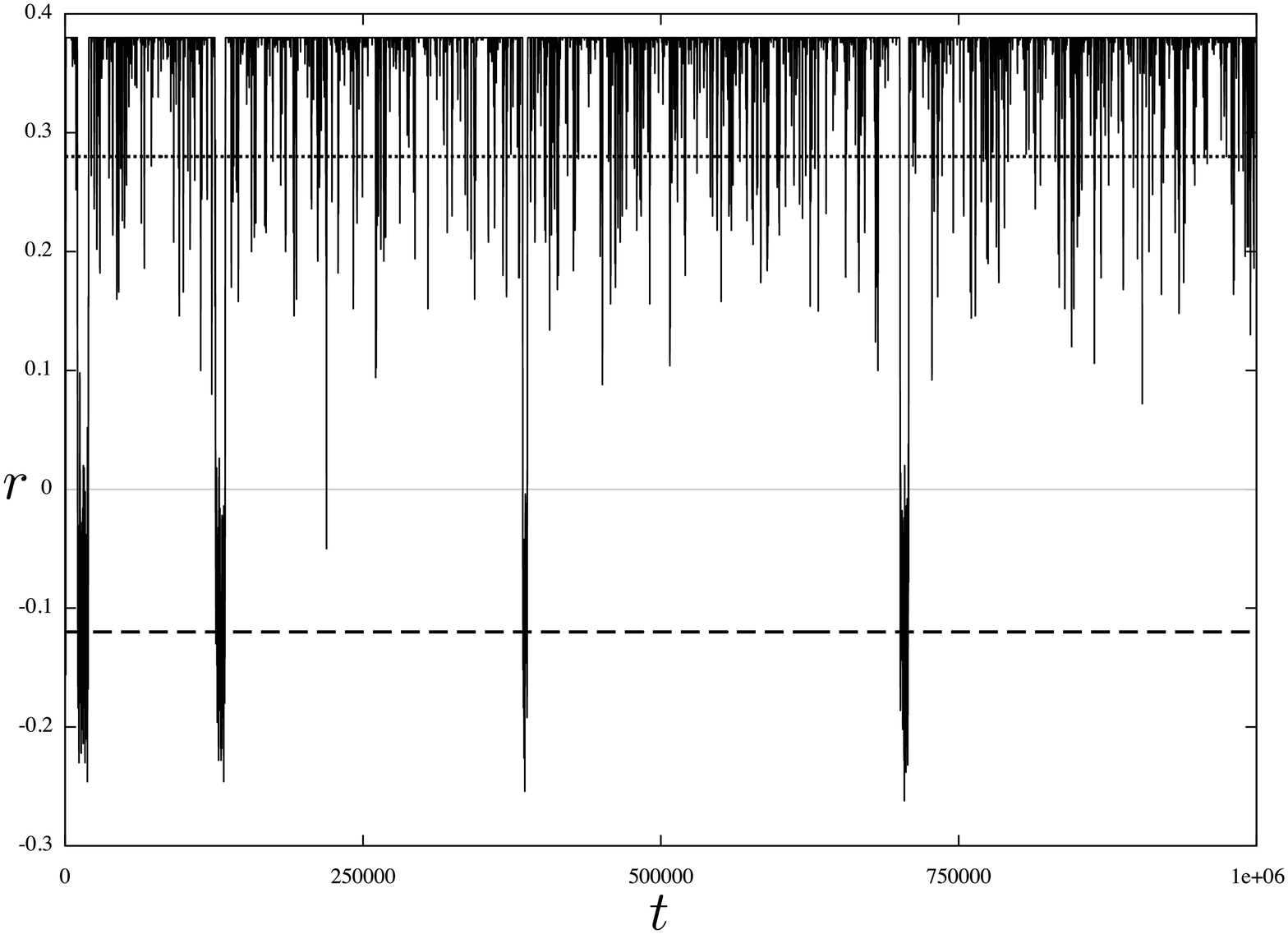}\\
(a)\\
 \includegraphics[width=7cm]{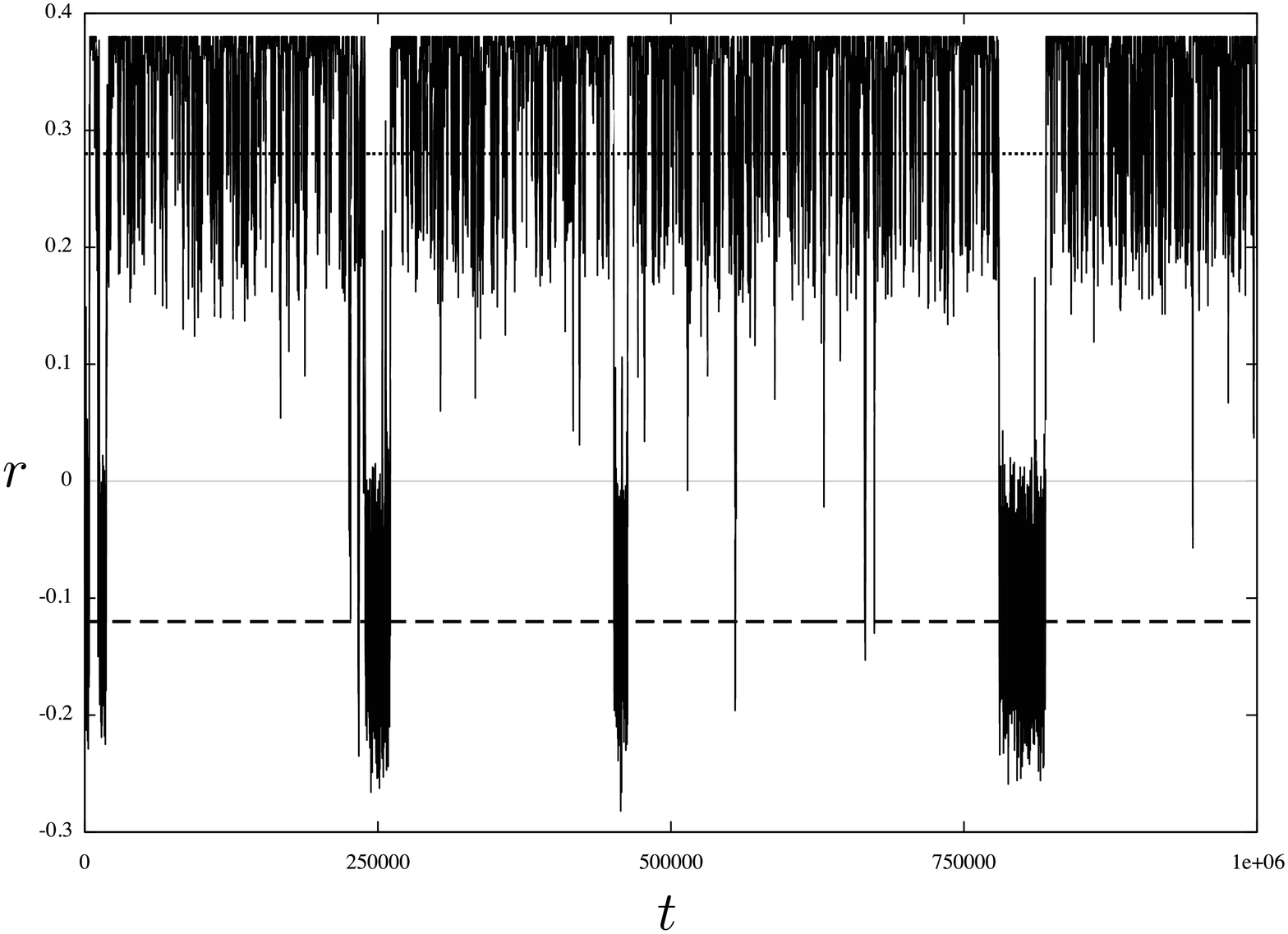}\\
(b)\\
 \includegraphics[width=7cm]{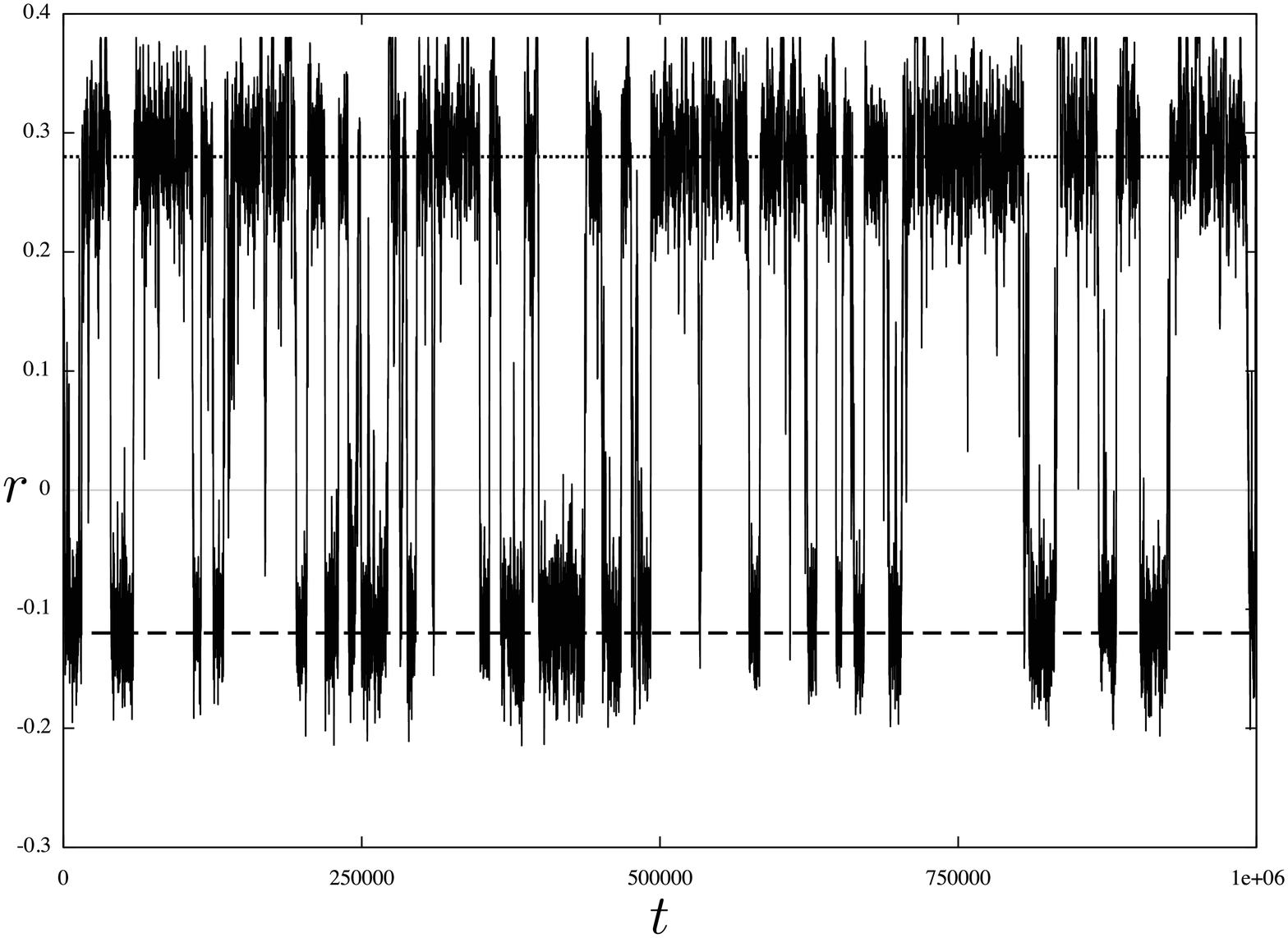}\\
(c)
\caption{Time series of $r$ in the system with mutation are plotted.
The payoff matrices of the games are set
$A=\left(\protect
\begin{smallmatrix} 0 & 0.2 \\ 0.8 & 0 \protect\end{smallmatrix}\right)$
and
$B=\left(\protect
\begin{smallmatrix} 0 & 0.3 \\ 0.7 & 0 \protect\end{smallmatrix}\right)$.
Thus, $p=0.2$ and $q=0.3$. Moreover, the selection intensity is $\omega=0.5$
and mutation rate is $\mu=5\times 10^{-5}$.
The number of individuals is $500$, $1,000$ and $2,000$ in (a), (b), and (c),
respectively.
The initial states are all set at
$\left(\protect
\begin{smallmatrix} x_{11} & x_{12} \\x_{21} & x_{22} \protect\end{smallmatrix}\right)
=\left(\protect
\begin{smallmatrix} pq & p(1-q) \\ (1-p)q & (1-p)(1-q) \protect\end{smallmatrix}\right)$.
The dashed lines represent $r^H(=0.28)$ and $r^L(=-0.12)$. The upper one is $r^H$ and the other is $r^L$.
}
\label{fig:r_mutation}
\end{center}
\end{figure}

To confirm the discussion in the last of the previous section,
let us introduce mutation into the system.
In each iteration of the process,
a randomly chosen individual replaces his/her strategy with another strategy
with a certain small probability $\mu$.
Since mutation makes the system ergodic, the system state shows round trips
along $L$ repeatedly. Therefore, we can observe the tendency of the motion.
In Fig. (\ref{fig:r_mutation}), the time series of $r$ in several population
sizes are plotted.
$r$ fluctuates around $r^H$ and $r^L$ and move back and forth between them repeatedly.
We can clearly observe that $r$ takes more time to fluctuate around $r^H$ than around $r^L$ in all plots.
As Eq. (\ref{eq:dotr}) suggests, this trend is observed more distinctly in a smaller population.

\subsection{Direction of flow at $(r,u,v)=(0,0,0)$}
\begin{figure}[htb]
\begin{center}
 \includegraphics[width=8cm]{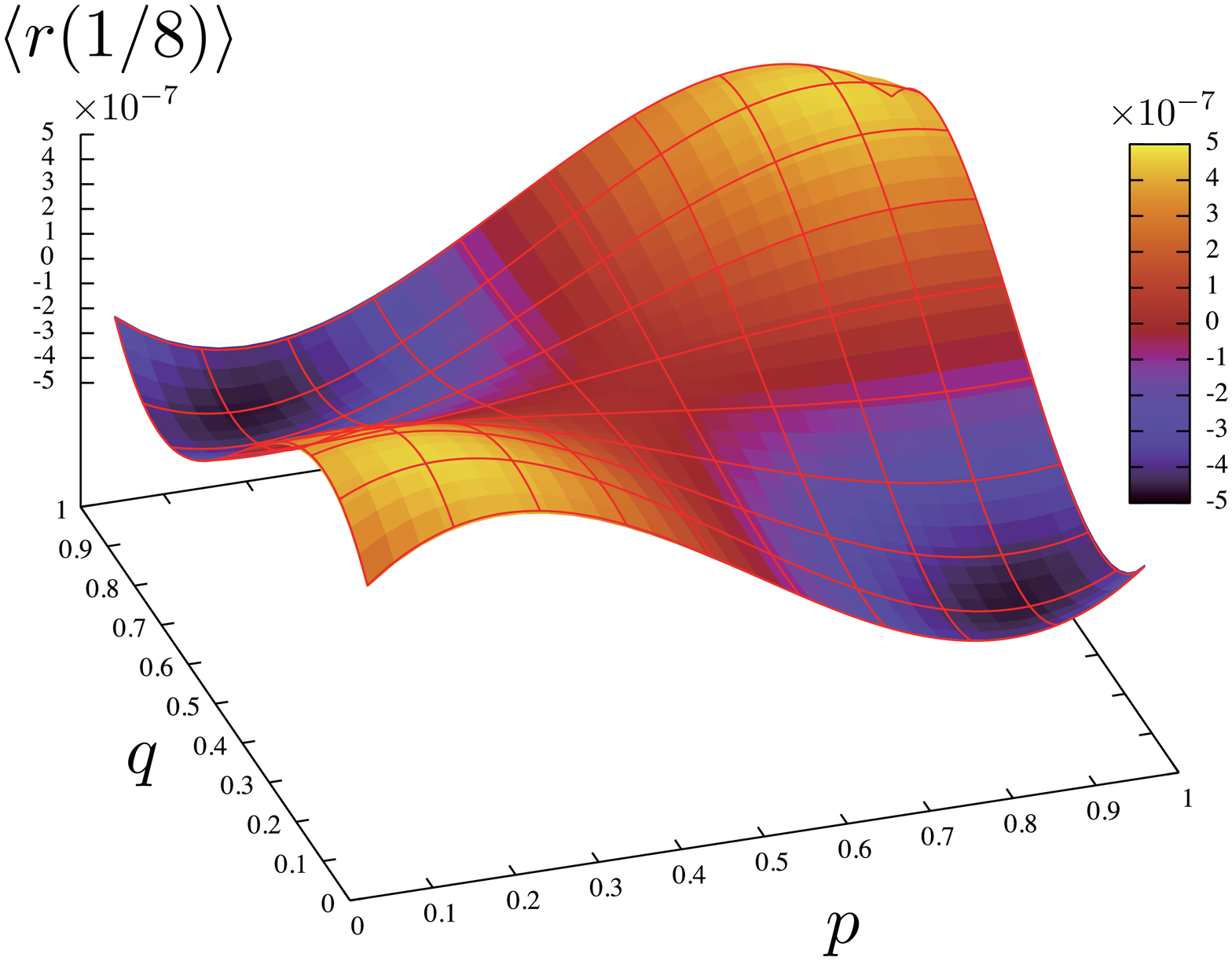}\\
\caption{
The number of players is $N=32^2=1024$.
The expected values of $r$ at $t=\frac{1}{8}$,
which start from $(r,u,v)=(0,0,0)$ at $t=0$, are plotted for various $(p,q)$.
$(p,q)$ is set at $(p,q)=\left(\frac{P}{32},\frac{Q}{32}\right),\;(P=1,\dots,31,\;Q=1,\dots,31)$.
The payoff matrices are given by
$A=\left(\protect
\begin{smallmatrix} 0 & p \\ 1-p & 0 \protect\end{smallmatrix}\right)$
and
$B=\left(\protect
\begin{smallmatrix} 0 & q \\ 1-q & 0 \protect\end{smallmatrix}\right)$ here.
%{\scriptsize $A = \begin{pmatrix} 0 & p \\ 1-p & 0 \end{pmatrix}$}
The initial state $\left(\protect\begin{smallmatrix} N_{11} & N_{12} \\ N_{21} & N_{22} \protect\end{smallmatrix}\right)$
($N_{ij}$ is the number of $(i,j)$-strategy players) corresponding to $(r,u,v)=(0,0,0)$
is $\left(\protect\begin{smallmatrix} PQ & P(32-Q) \\ (32-P) & (32-P)(32-Q) \protect\end{smallmatrix}\right)$.
The colored surface represents the results of the numerical simulation and
the red lines are plotted by Eq. (\ref{eq:r_1}).
Numerical simulation has been conducted as follows:
initially, a probability distribution on the system states
that concentrated on $\left(\protect\begin{smallmatrix} PQ & P(32-Q) \\ (32-P) & (32-P)(32-Q) \protect\end{smallmatrix}\right)$ is prepared
and updated $128\left(=N\times\frac{1}{8}\right)$ times with the transition probabilities given in Eq. (\ref{eq:P}); subsequently,
$\left<r(t=1/8)\right>$ is evaluated with the distribution at $t=\frac{1}{8}$.
}
\label{fig:r2d}
\end{center}
\end{figure}

Second, we calculate the expected values of $r$ at $t=\frac{1}{8}$ starting from
$(r,u,v)=(0,0,0)$ at $t=0$ for a set of various $(p,q)$.
Eq. (\ref{eq:dotr}) indicates that if both of $p$ and $q$ are larger or smaller than $\frac{1}{2}$,
$\left<r(t)\right>$ is positive, otherwise, $\left<r(t)\right>$ is negative.
A probability distribution, the initial distribution of which is
concentrated on $(r,u,v)=(0,0,0)$, is updated $N\times\frac{1}{8}$ times
with transition probabilities given by Eq. (\ref{eq:P}); subsequently,
$\left<r(t=1/8)\right>$ is evaluated with the distribution at $t=\frac{1}{8}$
and plotted in Fig. \ref{fig:r2d}.
As expected from Eq. (\ref{eq:dotr}),
Fig. \ref{fig:r2d} shows that when $p$ and $q$ are both larger or smaller than $\frac{1}{2}$,
$\left<r\right>$ is positive; otherwise, $\left<r\right>$ is negative.

To justify the framework of our analysis,
let us compare the values obtained by the numerical simulations
with analytically approximated values.
Since the period from $t=0$ to $t=\frac{1}{8}$ is too short
to use Eq. (\ref{eq:dotr}) as it is without any modifications,
we here derive the approximated value of $\left<r(t=1/8)\right>$
with the condition $(r(0), u(0), v(0)) = (0,0,0)$.
Since this initial state is a fixed point,
diffusional effect dominates the dynamics
when $t$ is small.
Thus,
the mean values of $u^2(t)$, $v^2(t)$, and $u(t)v(t)$ for $t\ll 1$
are simply obtained from
Eqs. (\ref{eq:huu})-(\ref{eq:huv}):
\begin{align}
&\left<\begin{pmatrix}u^2(t) & u(t)v(t) \\ u(t)v(t) & v^2(t)\end{pmatrix}\right>=
%\nonumber \\
%&\quad\quad\quad
\frac{1}{N} \begin{pmatrix}p(1-p) & 0 \\ 0 & q(1-q)\end{pmatrix}t+O(t^2).
\label{eq:Gamma2}
\end{align}
Additionally, the mean values of $r(t)u(t)$ and $r(t)v(t)$ for $t\ll 1$
are also obtained from Eqs. (\ref{eq:hru}) and (\ref{eq:hrv}):
\begin{align}
\left<\begin{pmatrix} r(t)u(t) \\ r(t)v(t)\end{pmatrix}\right>
= -\frac{1}{N}
\begin{pmatrix}p(1-p)(1-2q) \\  (1-2p)q(1-q)\end{pmatrix}t+O(t^2).
\label{eq:Gamma3}
\end{align}
From Eqs. (\ref{eq:Gamma2}), (\ref{eq:Gamma3}), and (\ref{eq:gr}),
$\dot{r}$ for $t\ll 1$ can be approximated by
\begin{align}
  \dot{r} &=-\frac{\omega t}{N}
\left\{\varPhi p(1-p)+\varPsi q(1-q)\right\}(1-2p)(1-2q)% \nonumber \\
%& \quad\quad\quad\quad
+O(t^2) +\frac{1}{\sqrt{N}} \zeta_r(t).
 \label{eq:dotr2}
\end{align}
$\left<r(t)\right>$ can be approximated by integrating Eq. (\ref{eq:dotr2}) as follows:
\begin{align}
\left<r(t)\right>
&= -\frac{\omega t^2}{2N} \left\{\varPhi p(1-p)+\varPsi q(1-q)\right\}(1-2p)(1-2q)
%\nonumber \\
%&\quad\quad\quad
+O(t^3).
\label{eq:r_1}
\end{align}
Fig. \ref{fig:r2d} shows
the approximation values calculated by Eq. (\ref{eq:r_1}).
This figure clearly indicates that
the values obtained by the numerical simulations are approximated effectively
by Eq. (\ref{eq:r_1}).

\section{Conclusion}
In this article,
we investigate a finite population with two games and
show that a finite population playing two games
tends to evolve toward a specific direction to form certain linkages
between the strategies of the two games.
We found that although the two games are not related,
a population tends to form a linkage between
the minor (major) strategies of the two games.

From the population genetics perspective,
this means that two loci,
which determine an individual's traits that
independently contribute to its fitness,
may have a stronger tendency to form a particular linkage disequilibrium
in smaller populations.
In future studies, more complicated situations, such as games that
have three or more strategies,
populations that play three or more games,
and diploid cases, could be investigated.

\bibliographystyle{model4-names}
\bibliography{mgd-bib}

\end{document}